\documentclass[aps,prb,reprint,superscriptaddress,longbibliography]{revtex4-2}
\usepackage{graphicx}
\usepackage{textcomp}
\usepackage{gensymb}
\usepackage{amsmath, amssymb}
\usepackage[colorlinks = true, citecolor = magenta]{hyperref}
\usepackage{braket}
\usepackage{natbib}
\usepackage[utf8]{inputenc}

\begin{document}

\title{A fluxonium qubit-based hybrid electromechanical system}

\author{Roson~Nongthombam}
\email{n.roson@iitg.ac.in}
\affiliation{Department of Physics, Indian Institute of Technology Guwahati, Guwahati-781039, (India)}

\author{Anshika~Ranjan}
\email{anshikar@iisc.ac.in} 
\affiliation{Department of Physics, Indian Institute of Science, Bangalore-560012 (India)}

\author{Amarendra~K.~Sarma}
\email{aksarma@iitg.ac.in}
\affiliation{Department of Physics, Indian Institute of Technology Guwahati, Guwahati-781039, (India)}

\author{Vibhor~Singh}
 \email{v.singh@iisc.ac.in} 
\affiliation{Department of Physics, Indian Institute of Science, Bangalore-560012 (India)}

\date{\today}

\begin{abstract}
Superconducting fluxonium qubits show a highly tunable 
energy-level structure, with transition frequencies 
spanning from a few MHz to few GHz. 
This range is well-aligned to the operational frequencies 
of highly coherent micro- and nano-mechanical resonators, 
making fluxonium an attractive candidate for hybrid electromechanical 
systems.
In this work, we theoretically investigate a flux-tunable 
electromechanical system consisting of a fluxonium qubit 
coupled to a suspended mechanical resonator. 
The coupling arises from the motion-induced modulation 
of magnetic flux through the fluxonium loop, enabling 
both transverse and longitudinal electromechanical 
interactions that are tunable via external magnetic fields.
By optimizing the design parameters of the fluxonium qubit, we demonstrate the feasibility of achieving strong resonant single-photon coupling near the flux-frustration point. We analyze the system dynamics across different coupling regimes, identifying signatures of electromagnetically induced transparency (EIT) in the longitudinal regime and mode splitting in the transverse regime.
Additionally, we show that ground-state preparation of 
both subsystems is possible through sideband cooling of 
the mechanical resonator.
These results suggest that a fluxonium-based hybrid electromechanical 
device could be a promising platform for studying macroscopic 
quantum phenomena and for applications in quantum technology.
\end{abstract}

\maketitle
\section{Introduction}

Superconducting qubits have become central to the development 
of quantum technologies with various implementations 
demonstrating high-fidelity gates, long coherence times, 
and scalable architectures \cite{krantz_quantum_2019,
kjaergaard2020superconducting,
mamgain_review_2023, BliasRevModPhys.93.025005, KochPhysRevA.76.042319}. 
Among the different realizations of the superconducting qubits, 
recently the fluxonium qubits have gained a lot of  attention due 
to its tunable nature of energy spectrum, large anharmonicity, 
and high coherence \cite{manucharyan2009fluxonium,earnest_realization_2018,PhysRevX.9.041041,PhysRevLett.130.267001}. 
The fluxonium qubit, formed by shunting a Josephson junction 
by a linear inductor, shows a highly anharmonic energy-level 
spectrum that can be tuned over a large frequency range, 
from a few megahertz to several gigahertz 
\cite{manucharyan2009fluxonium,earnest_realization_2018,zhang_universal_2021, G_Zhu_PhysRevB.87.024510, Jen_Koch_PhysRevLett.103.217004, PhysRevLett.130.267001}. 
This enables access to regimes of strong coupling and 
flux sensitivity that are difficult to achieve with transmon 
or flux qubits. 
Recent experiments have demonstrated coherence times 
approaching milliseconds \cite{earnest_realization_2018}, 
fast and high-fidelity gate operations \cite{zhang_universal_2021,ding2023high}, 
and strong coupling to microwave resonators \cite{kuzmin_quantum_2019,lee_strong_2023}, 
establishing fluxonium as a robust and versatile platform for 
quantum computing \cite{Feng_2022,nguyen2022blueprint}.
Its insensitivity to charge noise and operation near
flux sweet spots contribute to the noise resilience. 
In addition, its compatibility with flux-tunable designs 
makes it a promising candidate for integration into hybrid quantum
systems \cite{lee_strong_2023,najera-santos_high-sensitivity_2024,gerashchenko_probing_2025,YChuAppl.Phys.Lett, Clerk_2020, Xiang_Hybrid}.
%

Electromechanical systems coupled with superconducting 
qubits have emerged as interesting platforms 
for exploring quantum control of macroscopic systems, 
and enabling applications in quantum sensing, and transduction
 \cite{oconnell_quantum_2010,aspelmeyer_cavity_2014,barzanjeh_optomechanics_2022, Lauk_2020, Mirhosseini_2020}.
In microwave domain, coupling between the mechanical motion 
and the electromagentic modes has been realized by capacitive, 
inductive and piezoelectric interactions
\cite{oconnell_quantum_2010,teufel_sideband_2011,pirkkalainen_hybrid_2013,viennot_phonon-number-sensitive_2018,rodrigues_coupling_2019,schmidt_sideband-resolved_2020,bera_large_2021, Xue_2007}.
Bulk acoustic wave (BAW) and high-overtone bulk 
acoustic resonators (HBARs) have been quite successful 
demonstrating, 
strong dispersive and resonant coupling between transmon qubits,
phonon-number-resolving measurements, quantum-state transfer, and encoding the qubit
information in a mechanical mode 
\cite{Y_chu_2017,arrangoiz2019resolving,bienfait2019phonon,Yang_Yu_2024}. 
Most experimental realizations of hybrid-electromechanical device
thus far have used the Cooper-pair box and transmon qubit.
In this context, the fluxonium qubit can offer unique advantages 
for hybrid systems due to its large anharmonicity, high coherence, and
flux tunability of energy spectrum over a large range.
In particular, by designing the fluxonium qubit appropriately, 
a low frequency qubit, called ``heavy-fluxonium" can be 
realized \cite{zhang_universal_2021}.
Such qubits have transition frequency in the range of MHz which 
aligns naturally with many highly coherent micro and nanomechanical 
systems \cite{poot_mechanical_2012,aspelmeyer_cavity_2014}. 
In contrast to the earlier proposed flux-qubit–mechanical systems that operated 
in the GHz regime and relied on small, parametrically modulated 
coupling, 
the MHz-frequency fluxonium enables direct resonance with 
mechanical modes, giving access to coherent energy exchange 
and strong single-photon coupling in a naturally 
resonant regime \cite{Wang_cooling,Xue_2007,jaehne2008ground, LIAO2023108992}.
In addition, the flux tunability of fluxonium allows one to 
continuously change the nature of the electromechanical interaction—from 
predominantly longitudinal to predominantly transverse—within 
a single device, a level of in situ control that is not available 
in earlier flux-qubit or transmon-based mechanical hybrid systems.
In recent works, strong dispersive coupling between fluxonium 
and a mechanical resonator operating at 690 MHz \cite{lee_strong_2023}, 
and ac-charge mediated coupling to a 4~MHz SiN membrane \cite{gerashchenko_probing_2025} 
has been demonstrated. 
In this work, we present a detailed theoretical 
investigation of a hybrid system comprising a fluxonium qubit 
coupled to a mechanical resonator by a flux-mediated
coupling. 
The electromechanical coupling can be realized by embedding 
the mechanical resonator in the inductive loop of the 
fluxonium qubit. 
In presence of an in-plane magnetic field, such a scheme results
in tunable transverse and longitudinal couplings between the
``qubit" and the mechanical resonator.
Similar to the previous studies on flux-coupled electromechanical 
devices, such an approach results in magnetic field tunable 
single-photon coupling rates, $g_{\Phi}$. 
We derive explicit expressions for $g_{\Phi}$ and find that judicious tuning 
of the external flux enables us to realize the fluxonium qubit 
in the MHz and GHz transition frequency regime, allowing strong 
dispersive interaction with the mechanical mode in a single
experimental setup. 
We explored both the semi-classical, where the mechanical 
resonator remains in the thermal state, and the quantum nature of the 
hybrid system, where both the qubit and the mechanical resonator 
are cooled to their ground states. 
On one hand, when the coupling between the qubit and the mechanical 
resonator is longitudinal, we carry out semi-classical calculations, 
akin to the so-called electromagnetically induced transparency 
(EIT) phenomenon \cite{Marangos01031998,fleischhauer2005electromagnetically}.
On the other hand, by tuning the qubit at the flux sweet-spot, 
we achieve transverse coupling between the mechanical 
resonator and the qubit. 
This led us to the study of the mode-splitting phenomena. 
In order to observe quantum phenomena of the considered 
hybrid system, the mechanical resonator needs to be cooled 
to its motional ground state \cite{Rabl_cooling, jaehne2008ground, Wang_cooling}. 
It is done by tuning the qubit to the GHz frequency. 
Then bringing back the qubit to the MHz frequency, 
and transversely coupling it to the mechanical resonator, 
we have investigated the quantum phenomena such as 
Rabi oscillations and quantum entanglement between 
the qubit and the mechanical resonator. 
We organize the paper as follows: in Section~\ref{sec: hamiltonian}, 
we derive the hybrid Hamiltonian and compute the single-photon 
coupling rate, highlighting its dependence on magnetic field and 
mechanical parameters. Section~\ref{sec: semi-classic} presents a 
semi-classical framework. We show that under pump–probe drive, 
an off-sweet-spot bias yields a longitudinal regime with an EIT-type 
transparency window. On the contrary a half-flux bias realizes 
a transverse regime exhibiting mode splitting. 
In Section~\ref{sec:Quantum} we probe the quantum regime through 
a three‐step protocol: (i) sideband cooling the mechanical mode 
with a red‐detuned qubit drive, (ii) deriving an effective Master 
equation for the cooled resonator; and (iii) executing a rapid 
flux sweep to realize coherent single-phonon Rabi oscillations 
and entanglement. 
Finally, we conclude the discussion with 
future directions for tunable fluxonium–mechanical devices.

\begin{figure*}
\centering
\includegraphics[width=150mm]{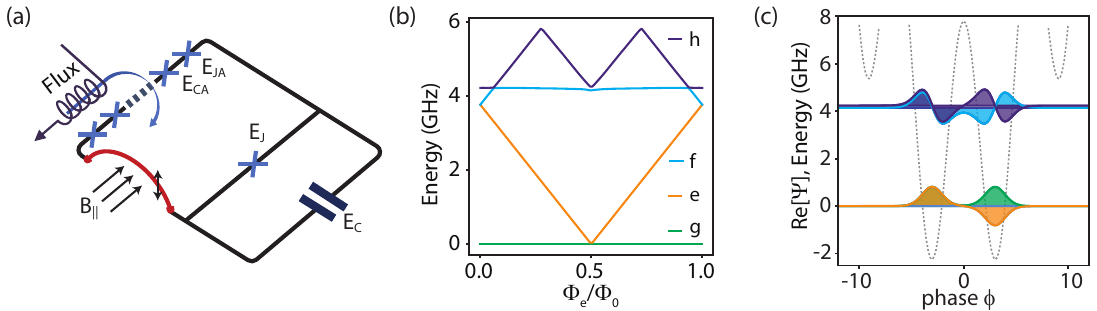}
\caption{ (a) Circuit diagram schematic of a fluxonium qubit with 
embedded mechanical resonator in the flux loop. The fluxonium qubit 
consists of a Josephson junction array, shunted by a single junction
and a capacitor. The flux-loop can be biased by an external flux
applied normal to the loop. In addition, an in-plane magnetic field
$B_{||}$ can be applied to realize electromechanical coupling. 
(b) Numerically calculated energy-eigenvalue spectrum of fluxonium 
qubit as external flux $\Phi_e$ is varied. All energy-levels are 
referenced to the ground state energy $E_g$. Real part of the wavefunctions 
and the corresponding potential at the flux point $\Phi_e = 0.5\Phi_0$ 
are shown in (c). The energy scale is the same as that in (b). 
The parameters used are $E_J/h = 5.5$~GHz, $E_C/h = 0.5$~GHz, and $E_L/h = 0.2$~GHz.}
\label{fig:fig1}
\end{figure*}

\section{Hamiltonian of the Hybrid System}
\label{sec: hamiltonian}

Consider a superconducting fluxonium qubit implemented using
a circuit that consists of a small area Josephson
junction with inductance \(L_J\), shunted by a capacitor
\(C_q\), and a superinductance of inductance \(L\).
The superinductance consists of an array of approximately
\(N = 300\) large-area Josephson junctions. 
Each junction in the array has the Josephson energy \(E_{JA}\) 
and the charging energy \(E_{CA}\), as shown in 
Fig.~\ref{fig:fig1}(a). 
We further assume that \(E_{JA} / E_{CA} \gg 1\), ensuring 
minimal charge dispersion at each junction of the array. 
This allows us to treat the array as an effective 
linear inductor.
The mechanical resonator is implemented within the 
fluxonium inductance loop itself by suspending one 
arm of the circuit to a length $l$ on the chip. 
The coupling between the fluxonium and the mechanical
resonator is achieved by applying an in-plane magnetic 
field $B_{||}$. 
For an out-of-plane mechanical mode, the displacement 
of the resonator $\hat{x}(t)$ results in the modulation
of the flux through the inductance loop by $\hat{\Phi}_m = B_{||} l\hat{x}(t)$.
Note that the fluxonium qubit can be flux-biased to $\Phi_e$ using
an independent perpendicular magentic field. 
The Hamiltonian of the fluxonium in the presence of 
the motionally induced flux can be written as,
\begin{equation}
\label{Fluxonium-mech1}
\hat{H} = 4E_C\hat{n}^2 - E_J \cos(\hat{\phi}) + \frac{E_L}{2} \left(\hat{\phi}-(\Phi_e + \hat{\Phi}_m)/\phi_0\right)^2,
\end{equation}
where $E_C = e^2/2C_q$, $E_J = \phi_0^2/L_J$, 
$E_L =\phi_0^2/L$, and \(\phi_0 = \Phi_0/2\pi = \hbar/2e\) is 
the reduced flux quantum.

Note that since the displacement $\hat{x}(t)$ of 
the resonator is time-dependent, it is convenient 
to place the external flux term in the qudratic term,
while $\Phi_e$ can be kept in the cosine term.
When $\hat{\Phi}_m(t)$ is added to the cosine term in Eq.~(1), 
a time derivative of $\hat{\Phi}_m(t)$ appears in the Hamiltonian. 
This can be eliminated by imposing an irrotational constraint 
on the flux variable. This is equivalent to adding the 
time-dependent flux term in the quadratic potential, 
rather than in the cosine potential 
(the last term in Eq.~(1))\cite{Xinyuan_You_2019}.
Expanding Eq. \ref{Fluxonium-mech1}, 
it is straightforward to obtain, 
\begin{eqnarray}
\label{Fluxonium-mech2}
\hat{H} &=& \hat{H}_q + \hat{H}_m+ \hat{H}_{i}   \nonumber \\
\hat{H} &=& \left(4E_C\hat{n}^2 -  E_J \cos (\hat{\phi} + \frac{\Phi_e}{\phi_0} ) + \frac{1}{2}E_L\hat{\phi}^2\right)  \nonumber \\ 
        &&+ \hbar\omega_m\hat{b}^\dagger\hat{b} \nonumber \\ 
        &&+ \hbar g_{\Phi}(\hat{b}+\hat{b}^\dagger)\hat{\phi}.
\end{eqnarray}
Here, $\hat{b}^\dagger$ and $\hat{b}$ are the creation and 
annihilation operators for the mechanical mode of 
frequency $\omega_m$.
The single-photon transverse coupling strength between the 
fluxonium and the resonator is given by $g_{\Phi}/2\pi = (E_L/h)(B_{||}lx_0/\phi_0)$, 
where $x_0 = \sqrt{\hbar/(2m\omega_m)}$ is the quantum zero-point 
displacement of the resonator. %
For a typical resonator of $(l\times w\times  t)= (40~\mu\text{m} \times 200~\text{nm} \times 28~\text{nm} )$, $m=0.75$~pg, $\omega_m/2\pi=6$~MHz, 
and for an applied field of $B_{||} = B_0$~mT and $E_L/h=0.2$~GHz, the 
coupling strength is $g_{\Phi}/2\pi=1049.8\times B_0$~Hz, where 
$B_0$ is the strength of the applied magnetic field 
in mT. 
We also note that the interaction also shifts
the mechanical mode frequency by an amount
$\hbar^2g_{\Phi}^2/(2E_Lm\,\omega_m x_0^2)$, which is
much smaller than the resonator frequency. 
Therefore, we have neglected this shift in Eq.~\ref{Fluxonium-mech2}.
The first three terms in Eq.~\eqref{Fluxonium-mech2} represent 
the bare fluxonium Hamiltonian. The fluxonium energy spectrum for $\Phi_{e}$ ranging from 0 to $\Phi_{o}$ is shown 
in Fig.~\ref{fig:fig1}(b). The parameters are chosen to 
realize a heavy fluxonium regime, using large shunting capacitors. 
As seen in the spectrum, the energy levels are highly anharmonic. 
Therefore, we can select any two levels at a given flux point 
and treat them as an effective two-level system.
Consider, for instance, the two-level subspace formed by the 
ground state \(|g\rangle\) and excited state \(|e\rangle\) at 
the flux point \(\Phi_e = 0.5\Phi_0\). 
At the half-flux point, the depths of the double-well potential 
are equal. This results in an equal wavefunction distribution 
across the two wells. Both the wavefunctions of the states 
\( \ket{g} \) and \( \ket{e} \) have equal amplitudes on 
both sides of the well; in other words, the wavefunction 
is not localized in one particular well. 
At this sweet spot, the fluxonium Hamiltonian is invariant under the inversion
$\hat{\phi} \rightarrow -\hat{\phi}$, and the eigenstates can therefore be chosen to have
definite parity. The ground state $|g\rangle$ is even (symmetric), while the first
excited state $|e\rangle$ is odd (antisymmetric) under this transformation see Fig\ref{fig:fig1}(c). Since the
phase operator $\hat{\phi}$ is odd under inversion, the diagonal matrix elements
$\langle g|\hat{\phi}|g\rangle$ and $\langle e|\hat{\phi}|e\rangle$ vanish by
symmetry, suppressing the longitudinal coupling, whereas the off-diagonal matrix
element $\langle g|\hat{\phi}|e\rangle$ is nonzero. As a result, the transverse
coupling matrix element becomes large, allowing single-photon transitions between
the $|g\rangle$ and $|e\rangle$ states at the half-flux point.
\par
\begin{figure}
\centering
\includegraphics[width=70mm]{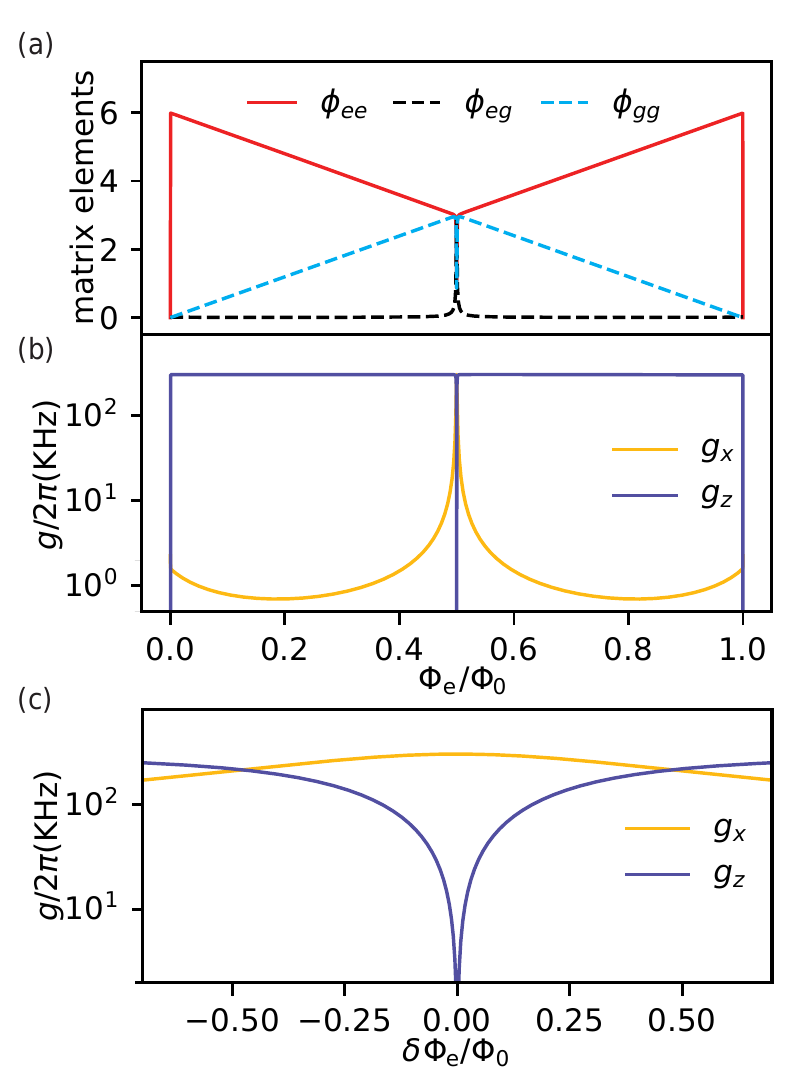}
\caption{(a) The absolute value of matrix elements of the 
phase operator $\hat{\phi}$, defined as: $\phi_{ij} = \langle i | \hat{\phi} | j \rangle$, 
where $\ket{i}$, and $\ket{j}$ are the fluxonium-eignestates. 
(b) Single-photon electromechanical coupling strengths $g_x =g_\Phi\phi_{eg}$ 
and $g_z= g_\Phi(\phi_{ee}-\phi_{gg})/2$ as external flux is varied. The coupling rate
$g_z$ vanishes at $\Phi_e/\Phi_0 = 0.5$, and $g_x$ acquires a maximum 
value. A detailed variation of single-photon electromechanical rates at half flux point 
is shown in panel (c) where $\delta\Phi_{e}/\Phi_0 =(\Phi_{e}/\Phi_0 - 0.5)*1000$. 
The coupling rates are calculated at an applied magnetic field $B_{||}$
of 100~mT and assuming quantum zero-point fluctuations of 
the mechanical resonator to be $x_0 = 43~\mathrm{fm}$.} 
\label{fig:fig2}
\end{figure}

%
In contrast, away from the half-flux point, the ground- and excited-state wavefunctions are mostly localized within a single well. As a result, the transverse coupling matrix element between them is very weak.
The strong coupling at (or very close) the half-flux point is further reflected in the symmetry of their wavefunctions: \(|g\rangle\) has even function, 
while \(|e\rangle\) is odd. Consequently, single-photon 
transitions between the two states are allowed due to 
the parity selection rule.
By identifying these two states as the logical states of a 
heavy-fluxonium qubit, its interaction with the mechanical 
mode can be captured in the Hamiltonian,

\begin{eqnarray}
\label{qubit_mech1}
\hat{H} & = & E_g |g\rangle\langle g| + E_e |e\rangle\langle e| + \hbar\omega_m\hat{b}^\dagger\hat{b} \nonumber \\ 
        && + \hbar g_{\Phi} \left[\phi_{gg} |g\rangle\langle g| + \phi_{ee} |e\rangle\langle e| \right](\hat{b}+\hat{b}^\dagger) \nonumber \\ 
        && + \hbar g_{\Phi} \left[\phi_{ge}|g\rangle\langle e| + \phi_{eg}|e\rangle\langle g| \right](\hat{b}+\hat{b}^\dagger),
\end{eqnarray}
where $E_{g,e}$ are the ground and excited level energies, respectively.
The matrix elements $\phi_{gg} = \langle g|\hat{\phi}|g\rangle$, 
$\phi_{ee} = \langle e|\hat{\phi}|e\rangle$ determine the longitudinal 
coupling, and the matrix elements $\phi_{ge} = \langle g|\hat{\phi}|e\rangle$ 
and $\phi_{eg} = \langle e|\hat{\phi}|g\rangle$ determine the transverse 
coupling to the mechanical mode.
A similar Hamiltonian to Eq.~\eqref{qubit_mech1} can be obtained
for the logical states \(|e\rangle\) and \(|f\rangle\) at the
flux point \(\Phi_e=0\). In this case also, transverse coupling between 
the logical states can be observed.
The matrix elements between the first two levels of the fluxonium 
as a function of external flux are shown in Fig.~\ref{fig:fig2}(a). 
At the half-flux point, the longitudinal coupling component 
vanishes, while the transverse component dominates. 
In the vicinity of the half-flux point, a non-zero longitudinal 
component also appears Fig.~\ref{fig:fig2}(b). The coupling rates are calculated following the relation $g_{\Phi}/2\pi = (E_L/h)(B_{||}lx_0/\phi_0)$ for a typical resonator of $l=40\mu$m, $m=0.75$~pg, $\omega_m/2\pi=6$~MHz, $E_L/h=0.2$~GHz at an applied magnetic field $B_{||}$
of 60~mT and assuming quantum zero-point fluctuations of 
the mechanical resonator to be $x_0 = 43~\mathrm{fm}$. The mechanical and circuit parameters used here are well within current experimental
capabilities \cite{zhang_universal_2021,bera_large_2021,gerashchenko_probing_2025,PhysRevLett.130.267001,pirkkalainen_hybrid_2013,rodrigues_coupling_2019,teufel_sideband_2011,viennot_phonon-number-sensitive_2018,schmidt_sideband-resolved_2020}.

\section{Semi-classical treatment}
\label{sec: semi-classic}

After establishing that both the longitudinal and the transverse 
couplings can be realized in this setup, we now present the semi-classical 
treatment of the Hamiltonian in Eq.~\ref{qubit_mech1} at two different flux 
operating points where either longitudinal or transverse coupling dominates.
The fluxonium qubit energy spectrum shown in Fig. \ref{fig:fig1}(b) 
exhibits both GHz and MHz transition frequencies. 
Due to large tunablity of the qubit-frequency, $g_z$ dominates over
the $g_x$ away from the sweet-spots of \(\Phi_e/\Phi_{o} = 0,\text{and}~0.5\).  
Particularly, at the flux sweet-spots, $g_z$ vanishes as the qubit 
frequency become first order flux in-sensitive, and $g_x$ becomes
the dominant coupling mechanism. 
Moreover, at an external flux point of \(\Phi_e = 0.5\Phi_{o}\), the 
transition frequency can be made quite small 4~MHz, which could be made
near resonant with the mechanical mode frequency. We next discuss these two 
cases separately.

\subsubsection*{Longitudinal coupling regime: ($g_z \gg g_x$)}

Let's consider the flux point \( \Phi_e = 0.3\Phi_{o}\). 
At this point, the transverse elements \(\phi_{eg}=\phi_{ge}=0.008\) 
are much smaller compared to the longitudinal matrix 
elements \( \phi_{gg} =-1.801\) and \(\phi_{ee}=4.2 \). 
Therefore, by ignoring the transverse elements and identifying
$\omega_q = (E_e-E_g)/\hbar$, the 
Hamiltonian in Eq. \ref{qubit_mech1} can be written 
in terms of the Pauli operators as,
\begin{eqnarray}
\label{longitudinal}
    \hat{H}_G &=& \frac{\hbar\omega_q}{2}\hat{\sigma}_z + \hbar\omega_m\hat{b}^\dagger\hat{b} + \frac{\hbar}{2} g_{\Phi} \, \phi_{gg} (1-\hat{\sigma}_z ) (\hat{b}+\hat{b}^\dagger) \nonumber\\ &&
    + \frac{\hbar}{2} g_{\Phi}\, \phi_{ee} (1 +\hat{\sigma}_z ) (\hat{b}+\hat{b}^\dagger).
\end{eqnarray}

For a MHz frequency mechanical resonator, the qubit transition frequency 
is highly detuned from the resonator frequency. As a result, the interaction 
between them is minuscule compared to the qubit's frequency scale. 
Such interaction can be parametrically enhanced by adding a drive tone
to the qubit detuned approximately by the mechanical resonant frequency. 
Along with a probe signal, the total Hamiltonian of the system using 
Eq.~\ref{longitudinal} can be reduced to:
\begin{eqnarray}
\label{longitudinal1}
    \hat{H}_{GD} &=& -\frac{\hbar\Delta_q}{2}\hat{\sigma}_z +  \hbar\omega_m\hat{b}^\dagger\hat{b} + \frac{\hbar}{2} g_{\Phi} \, \phi_{gg} (1-\hat{\sigma}_z ) (\hat{b}+\hat{b}^\dagger) \nonumber\\ &&
    +\, \frac{\hbar}{2} g_{\Phi}\, \phi_{ee} (1 +\hat{\sigma}_z ) (\hat{b}+\hat{b}^\dagger) + \hbar\,\epsilon_d(\hat{\sigma}_- + \hat{\sigma}_+) \nonumber \\ &&
    +\, \hbar\,\epsilon_p(\hat{\sigma}_- e^{i\delta_pt} + \hat{\sigma}_+e^{-i\delta_pt}),
\end{eqnarray}
where \( \Delta_q = \omega_d - \omega_q \) and \( \delta_p = \omega_p - \omega_d \) 
are the drive tone and probe tone detunings. 
The drive and probe tones have their amplitude as \( \epsilon_d \) and \( \epsilon_p \) , respectively. 
The rotating wave approximation is applied to the drive terms, 
assuming that \( \omega_d, \omega_p \gg \epsilon_d, \epsilon_p \).
Since the resonator is in a thermal state due to its low frequency, 
we treat the hybrid system semi-classically and evolve the dynamics 
of Eq.~\ref{longitudinal1} by applying the mean-field approximation. 
Taking the ansatz 
\( \langle\hat{O}\rangle = \langle\hat{O}_0\rangle + \langle\hat{O}_-\rangle e^{-i \delta_p t} + \langle\hat{O}_+\rangle e^{i \delta_p t} \) for the semi-classical variables, the spectrum for 
\( \langle\sigma_-^- \rangle\) is plotted in Fig.~\ref{classical_mode split}(a), 
where we include qubit dissipation through the energy relaxation rate
$\Gamma = 1/T_1$ of 2~MHz and a pure dephasing rate $\Gamma_\phi = 1/T_\phi$ 
of 0.9~MHz, which are used to compute the steady-state qubit response
(refer appendix \ref{appendix A}). 
As seen in the figure, a dip appears in the spectrum when the frequency 
difference between the probe and the detuning drive matches the resonator 
frequency, signaling the onset of electromagnetically induced transparency 
(EIT). This fit can be used to deduce 
the frequency of the resonator and the coupling strength.

\subsubsection*{Transverse coupling regime: ($g_z \ll g_x$)}
\begin{figure}
\centering     
\includegraphics[width=85mm]{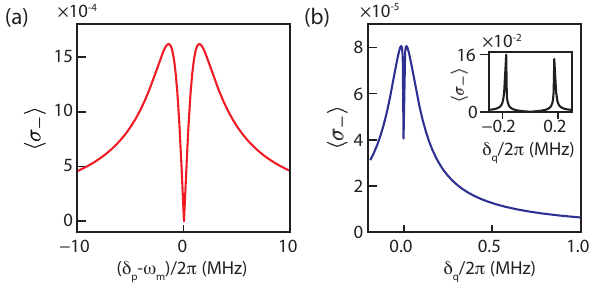}
\caption{(a) Spectrum of the fluxonium qubit (absolute value of the real part of amplitude $\langle\hat{\sigma}_-^-\rangle$) obtained using the pump-probe method
while operating in $g_z\gg g_x$ limit at $\Phi_e/\Phi_0=0.3$. 
Parameters used: Qubit energy relaxation rate of $\Gamma/2\pi=2$~MHz, 
pure dephasing rate of $\Gamma_{\Phi}/2\pi=0.9$~MHz, 
$\epsilon_p=\epsilon_d=5$~kHz, and thermal occupation of $n_q=0$.
(b) Mode splitting of the qubit–mechanical state. In the absence of qubit thermal occupation, the mode splitting is clearly observed (inset). However, when the qubit has a high thermal occupation, the splitting becomes indistinguishable. Both the mechanical resonator and qubit transition frequencies are 4 MHz. Note that the amplitude of the modes also decreases with increasing thermal occupation.
Parameters used: $ \Gamma/2\pi=\Gamma_{\Phi}/2\pi=1$~kHz, 
$\epsilon_p'=1$~kHz, $n_q=78$, $\epsilon'_p=1$~kHz. 
Common parameters used in both the plots:
$\omega_m/2\pi=4$~MHz and energy decay rate $\gamma_m/2\pi=5$~Hz, 
and $g_{\Phi}/2\pi=60$~KHz.} 
\label{classical_mode split}
\end{figure}

At the flux point \(\Phi_e = 0.5\Phi_0\), a significant transverse 
coupling between the qubit and the resonator can be achieved. 
The transverse and longitudinal matrix elements at this flux point 
are shown in Fig.~\ref{fig:fig2}(c). 
We note that there is no longitudinal component at this flux point. 
However, the longitudinal contribution becomes comparable to the 
transverse one in a small vicinity around the half-flux point. 
At exactly half flux, the Hamiltonian in Eq.~\eqref{qubit_mech1} 
can be written as:
\begin{eqnarray}
\label{transverse}
    \hat{H}_{M} = \frac{\hbar\omega_q}{2}\hat{\sigma}_z + \hbar\omega_m\hat{b}^\dagger\hat{b} + \hbar g_{\Phi} \, \phi_{eg} (\hat{\sigma}_- + \hat{\sigma}_+ ) (\hat{b}+\hat{b}^\dagger).
\end{eqnarray}
Here, the interaction strength between the qubit and the mechanical mode is only one order of magnitude smaller than their transition frequencies.
Therefore, we can directly apply the probe drive and observe 
the interaction. Assuming the probe amplitude \( e'_p \) and
coupling strength \( g_x = g_{\Phi}\phi_{eg}\) to be much 
less than the probe frequency, we have
\begin{eqnarray}
\label{transverse_probe}
    \hat{H}_{MD} &=& \frac{\hbar\delta_q}{2}\hat{\sigma}_z + \hbar\delta_m\hat{b}^\dagger\hat{b} + \hbar g_x (\hat{\sigma}_-\hat{b}^\dagger + \hat{b}\hat{\sigma}_+ ) \nonumber\\&& + \hbar e'_p (\hat{\sigma}_- + \hat{\sigma}_+ ),
\end{eqnarray}
where $\delta_q = \omega_q-\omega_p$ ($\delta_m = \omega_m-\omega_p$) is 
the detuning between the qubit (mechanical) frequency and the 
probe frequency $\omega_p$.
Initially, both the qubit and the mechanical resonator are
in thermal equilibrium with their baths. 
The qubit can be brought to the ground state by performing 
a reset protocol \cite{zhang_universal_2021}. This involves 
dispersively coupling the qubit with a cavity and selectively 
driving the qubit transition. 
We can then use the semiclassical approach to evolve the system
and obtain the spectrum of the qubit in steady state, as shown
in Fig.~\ref{classical_mode split}(b). 
Since the interaction is transverse, mode splitting should be 
observed at \( \omega_p = \omega_m \). But, due to the presence 
of thermal effects (with the qubit thermal phonon \( n_{\text{th}} = 78 \)), 
the spacing between the split modes is highly suppressed. 
To compare the decrease in the splitting, the qubit spectrum at 
zero thermal photons is also included in the figure. 
Notice that both the splitting and the amplitude of 
the spectrum decreases due to the presence of the thermal phonon. Moreover,when the qubit is detuned from the mechanical resonance, 
the splitting becomes asymmetric, and one peak dominates over the other.
This behavior resembles the dispersive shift of the qubit frequency, 
or equivalently the mechanical frequency, observed in a typical 
Jaynes--Cummings interaction in the dispersive limit.

\section{Quantum behavior}
\label{sec:Quantum}

To operate the hybrid system at the quantum level, 
both the qubit and the mechanical resonator must 
be cooled to their ground states.
At low temperatures, the qubit can be initialized to its 
ground state by tuning its transition frequency to 
GHz range. However, the mechanical mode
resonator with lower resonant frequency must be further 
cooled to reach the ground state using sideband cooling
techniques \cite{wilson-rae_laser_2004,marquardt_quantum_2007}. 
Here, the qubit acts as an effective bath for 
the mechanical resonator, enabling effective cooling or 
heating of the resonator. When the qubit is driven 
with a red-detuned tone, it effectively acts as a 
cooling bath for the resonator. 
In contrast, a blue-detuned drive turns the qubit into 
a heating bath for the resonator. 
An important assumption here is that the qubit reaches its 
steady state much faster than the resonator, allowing for 
the adiabatic elimination of the qubit dynamics. 
This adiabatic elimination relies on two standard conditions 
for the separation of timescales, namely $g_z \ll \Gamma$ and $n_m \gamma_m \ll \Gamma$, 
where $g_z$ is the longitudinal coupling rate, $\Gamma$ is 
the qubit energy relaxation rate, $\gamma_m$ is the mechanical 
damping rate, and $n_m$ is the thermal phonon occupation 
of the resonator. 
Since $g_z \propto B_0$, the applied in-plane magnetic field 
can always be reduced to satisfy the first condition, 
even for relatively small decay rates spanning tens to 
hundreds of kilohertz~\cite{lee_strong_2023,gerashchenko_probing_2025,zhang_universal_2021}.
On the mechanical side, we use $\gamma_m/2\pi = 5~\mathrm{Hz}$ 
and an initial occupancy $n_m \approx 50$, so that $n_m \gamma_m$ 
remains orders of magnitude smaller than $\Gamma$.
It is worth noting that the total energy relaxation rate $\Gamma$ depends 
on multiple physical mechanisms beyond the bare qubit decay rate. When the 
qubit frequency is tuned away from the sweet-spot for resetting—shifting 
from MHz to GHz frequencies—enhanced dielectric loss at higher frequencies, 
quasi-particle dissipation across the small-area junction, and Purcell 
damping from the readout resonator collectively reduce the relaxation time 
$T_1$ by orders of magnitude. This frequency-dependent enhancement of 
dissipation has been experimentally observed in heavy-fluxonium devices 
\cite{pop2014coherent}, ensuring that the hierarchy $g_z \ll \Gamma$ is 
naturally satisfied during the cooling protocol.

We first write down the driven qubit-mechanical Hamiltonian 
from Eq.~\eqref{longitudinal}. 
\begin{eqnarray}
\label{cooling}
    \hat{H}_{C} &=& -\frac{\hbar\delta_q}{2}\hat{\sigma}_z +  \hbar\omega_m\hat{b}^\dagger\hat{b} + \hbar g_z (\hat{b}+\hat{b}^\dagger)\hat{\sigma}_z   \nonumber\\ &&
     + \hbar\,\epsilon_{r}(\hat{\sigma}_- + \hat{\sigma}_+), 
\end{eqnarray}
where $g_z = g_{\Phi}(\phi_{ee} - \phi_{gg})/2$. 
The detuning is given by $\delta_q=\omega_r-\omega_q$, 
and the strength of the drive is denoted by $\epsilon_r$.
To perform the adiabatic elimination of the qubit, 
we first write down the Lindblad master equation for 
the coupled qubit-mechanical system using the above Hamiltonian. 
We then trace out the qubit degrees of freedom, resulting 
in an effective master equation for the mechanical resonator\cite{jaehne2008ground}.
\begin{eqnarray}
    \label{master equation}
    \dot{\hat{\rho}}_m &=& -i[(\omega_m + \delta\omega_m)\hat{b}^\dagger\hat{b}, \hat{\rho}_m] + [\Gamma_q^- + \gamma_m(n_{m} + 1)] D[\hat{b}]\hat{\rho}_m
    \nonumber \\ &&+ [\Gamma_q^+ + \gamma_m n_{m}] D[\hat{b}^\dagger]\hat{\rho}_m,
\end{eqnarray}
where, 
\begin{eqnarray}
    \label{Gamma}
    \Gamma_q^- &=& g_z^2 \, \Re[S(\omega_m)] \nonumber \\
    \Gamma_q^+ &=& g_z^2 \, \Re[S(-\omega_m)] \nonumber\\ 
    \delta\omega_m &=& g_z^2 [\Im[S(\omega_m)] - \Im[S(-\omega_m)]].
\end{eqnarray}
Here, $\mathcal{D}[\hat{O}]\rho = \hat{O}\rho\hat{O}^\dagger - 
\frac{1}{2}\{\hat{O}^\dagger\hat{O}, \rho\}$ is the Lindblad dissipator 
operator, which describes the effect of dissipation from the decay of 
operator $\hat{O}$. 
These quantities $\hat{\rho}_m$, $n_m$, and $\gamma_m$ denote the 
density operator, mean-thermal occupation, and 
decay rate of the mechanical resonator, respectively.

\begin{figure}
\centering
\includegraphics[width=80mm]{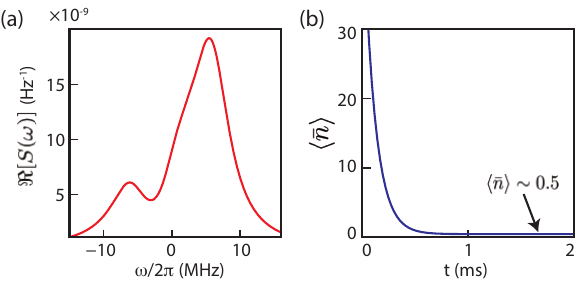}
\caption{(a) Real part of the one-sided power spectral density of the qubit in the steady state. Two peaks are observed in the vicinity of $\omega/2\pi = \pm\,\omega_m/2\pi = 6~\text{MHz}$. The spectral values at $\omega = \pm \omega_m$ correspond to the qubit-induced decay rates, $\Gamma_q^\pm$. The difference in these spectral values at $\omega = \pm \omega_m$ determines the effective cooling rate of the resonator.
(b) Mean phonon occupation of the mechanical resonator. The resonator is initially cooled to a few phonons (30 in this case) by connecting it to a millikelvin bath realized with a dilution refrigerator. Then, by coupling the resonator to the qubit acting as a cold bath, it is further cooled to its ground state.
Parameters used are:$\Phi_e=0.3$,
$\gamma_m/2\pi=5$~Hz, $\Gamma/2\pi=2$~MHz, $\Gamma_\Phi/2\pi=0.9$~MHz, 
$\epsilon_r=0.89\,\omega_m$, and $\delta_q=-\sqrt{\omega_m^2-\epsilon_r^2}$}
\label{fig:qubit_cooling}
\end{figure}

The quantity $S(\omega) = \int_{0}^{\infty} d\tau\, e^{i\omega \tau} \langle \delta\hat{\sigma}_z(\tau)\, \delta\hat{\sigma}_z(0) \rangle_{\text{ss}}$ is the single-sided power 
spectral density of the qubit in steady state, 
where 
\(\delta\hat{\sigma}_z = \hat{\sigma}_z - \langle \hat{\sigma}_z \rangle\). 
This spectrum depends on both the strength of the red-detuned 
drive applied to the qubit and the qubit’s intrinsic decay rates.
The qubit-induced decay rate \(\Gamma_q^-\) leads to cooling 
of the mechanical resonator, while \(\Gamma_q^+\) contributes to heating. 
The real part of \(S(\omega)\), which determines these decay rates, 
is shown in Fig.~\ref{fig:qubit_cooling}(a). 
When the qubit is driven at a frequency below its transition 
frequency (\(\delta_q < 0\)), the cooling rate dominates 
over the heating rate, as illustrated in Fig.~\ref{fig:qubit_cooling}(b). 
The optimized cooling is achieved for $\epsilon_r=0.89\,\omega_m$ 
and $\delta_q=-\sqrt{\omega_m^2-\epsilon_r^2}$ \cite{jaehne2008ground,jahne2009cavity}.
This cooling mechanism relies on the qubit absorbing 
phonons from the resonator and becoming excited, thereby 
extracting energy from the mechanical mode. 
Once the mechanical resonator is sufficiently cooled, 
the drive on the qubit is switched off, allowing the qubit 
to relax back to its ground state. 

\begin{figure}[ht]
\centering
\includegraphics[width=75mm]{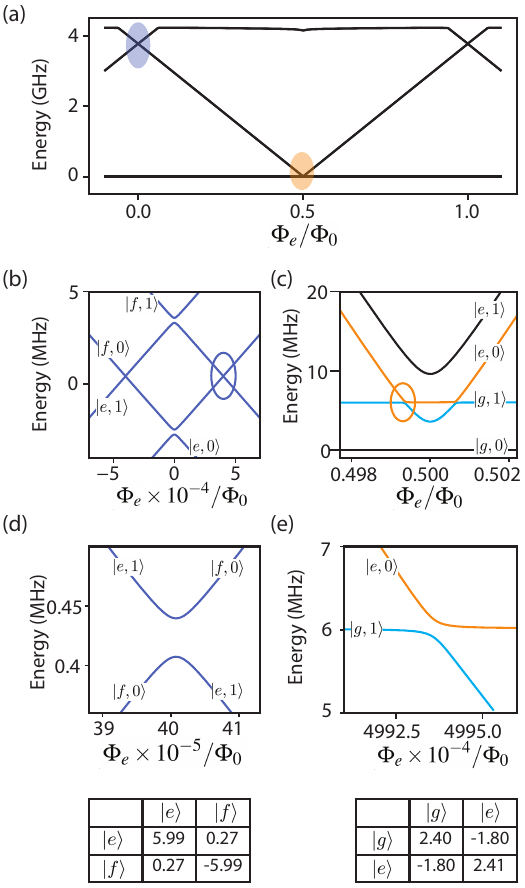}
\caption{(a) Energy spectrum of the composite system
as external flux $\Phi_e$ is varied. Due to large difference 
between energy scales of qubit and mechanical 
mode, the granularity in the spectrum is not clearly visible
at this energy scale. Panel (b) and (c) shows the detailed view 
of the avoided crossing near flux symmetric points $\Phi_e/\Phi_0$~=~0, 
and 0.5.
Multiple level-crossings with single phonon state can be clearly 
seen. For clarity, the y-axis values shown in panel (b) has been
offset by 3.772~GHz.
Panel (d) and (e) shown a further zoomed-in view of the 
energy level crossing showing the avoided crossing arising from 
the hybridization with the single-phonon state.
Bottom Tables: Matrix elements of the operator $\hat{\phi}$ at 
the symmetric points(left: at $\Phi_e/\Phi_0$ = 0.0004, 
right: at $\Phi_e/\Phi_0$ = 0.49938).
The off-diagonal elements of the matrix determine the 
amount of energy splitting shown in (d) and (e).}
\label{fig:qubit_mech_spectrum}
\end{figure}

Having cooled down the mechanical resonator by utilizing the 
qubit's cold bath and the strong longitudinal coupling between 
the qubit and the resonator, we can observe coherent interactions 
between them at the single-phonon excitation level. 
To demonstrate this, we plot the energy spectrum of the 
composite system—namely, the fluxonium qubit and the mechanical 
resonator with a single phonon excitation, as shown in 
Fig.~\ref{fig:qubit_mech_spectrum}. 
As seen in the figure, level-crossings
occur at two pairs of symmetric flux points -- one near 
\( \Phi_e = 0\), and another near the flux-frustration point
\( \Phi_e = 0.5\Phi_0 \).
Near \( \Phi_e = 0\), the first excited state $\ket{e}$, and
the second excited state $\ket{f}$ nearly ``touch" at the 
zero-flux point.
As a result, a single excitation in the mechanical mode shifts 
this spectrum approximately $\omega_m$, and the electromechanical 
coupling results in the levels repulsion in the energy spectrum 
of the composite system as shown in Fig.~\ref{fig:qubit_mech_spectrum}(a) 
to (e).
Similar avoided crossings are observed near \( \Phi_e = 0.5 \Phi_0\) when 
the fluxonium states $\ket{g}$, and $\ket{e}$ mix with the single-phonon
state.
In should be noted that these symmetry points repeat periodically 
as the external flux \( \Phi_e \) is varied. In addition, at each 
of these points, avoided crossings can occur due to higher 
excitations in the mechanical mode. 
The amplitude of these crossings depends on the phonon number, 
reflecting the quantum nature of the qubit–mechanical interaction. 
Additional details, including the dispersive shift at higher 
excitations, are given in Appendix \ref{Appendix B}

We note that the coherent single-excitation exchange occurs in a narrow flux window
close to the half-flux sweet spot, as evident from Fig.~5(d,e). At
$\Phi_e = 0.5\Phi_0$, the fluxonium transition frequency is first-order insensitive
to flux fluctuations, i.e., $\partial\omega_q/\partial\Phi_e = 0$, and residual
flux noise enters only through the second-order curvature
$\partial^2\omega_q/\partial\Phi_e^2$.
Following the curvature in Fig.~2(c), a flux deviation 
$\delta\Phi_e \sim 10^{-5}\Phi_0$ corresponds to a frequency shift that is negligible
compared to the electromechanical coupling strength
$g_\Phi/2\pi \sim 60~\mathrm{kHz}$.
Furthermore, since the excitation exchange occurs on a timescale set by 
the transverse coupling, which is much shorter than the 
correlation time of low-frequency $1/f$ flux noise, flux 
fluctuations are effectively quasi-static during the interaction. 
Consequently, flux noise primarily produces a small static 
detuning rather than dynamical dephasing, and does not wash 
out the coherent single-excitation exchange.
This also ensures that neither the qubit nor the resonator 
significantly relax to its thermal state during the 
transition. 
A natural concern at this stage is whether operation of the fluxonium qubit at
a low transition frequency $\omega_q/2\pi \simeq 4~\mathrm{MHz}$ near the
half-flux sweet spot $\Phi_e = 0.5\Phi_0$ makes the protocol overly sensitive
to flux noise. At this bias point the qubit frequency is only first-order
insensitive to flux fluctuations, so in principle residual second-order
coupling to flux noise could limit coherence. However, experiments on
heavy-fluxonium devices operated at similar sweet spots and frequencies show
that this residual sensitivity does not set the dominant decoherence scale:
coherence times $T_2^* > 1~\mathrm{ms}$ have been reported for low-frequency
fluxonium qubits \cite{nguyen2022blueprint,PhysRevLett.130.267001}, and coherence in excess of
$300~\mu\mathrm{s}$ has been demonstrated for fluxoniums in the tens-of-MHz
range \cite{zhang_universal_2021}. 
Therefore, existing experimental evidence and
the timescales of our protocol indicate that second-order flux noise at the
4~MHz operating point should not prevent the observation of the coherent Rabi
oscillations.

To observe the coherent interaction between the qubit and the resonator, we first excite the qubit to the state \(|e,0\rangle\) at the flux point \(\Phi_e = 0.3\,\Phi_0\), and then rapidly sweep the flux to bring the qubit into resonance with the mechanical resonator. 
The sweep is designed such that the system remains predominantly in the state \(|e,0\rangle\) upon reaching \(\Phi_e = 0.49936\,\Phi_0\), which corresponds to the avoided crossing.
The sweep is performed in less than \(1\,\mu\text{s}\), ensuring that the fluxonium does not decay to the thermal state before reaching the avoided crossing. 
However, for such fast sweeps, Landau--Zener transitions can occur between fluxonium states, which reduces the population of the \(|e,0\rangle\) state prepared at the avoided crossing. 
Nevertheless, a significant amount of coherent exchange of excitation can still be observed for sweep times of about \(1\,\mu\text{s}\). 
The effect of Landau--Zener transitions on the Rabi oscillations is discussed in Appendix~\ref{appendix:C}.

\subsubsection*{Protocol for generating qubit–mechanical entanglement}

We now outline a concrete experimental sequence to create and observe entangled states between the fluxonium qubit and the mechanical resonator, starting from a thermal mechanical state.

\begin{enumerate}
    \item \textbf{Initialization.} The qubit is flux-biased at \(\Phi_e = 0.3\Phi_0\), where its transition frequency is \(\omega_q/2\pi \approx 2\)~GHz. At a base temperature of \(\sim15\)~mK, the qubit thermal occupation is negligible (\(n_q \approx 0\)). The mechanical resonator (frequency \(\omega_m/2\pi = 6\)~MHz) is in a thermal state with mean phonon number \(n_m \approx 50\), determined by its coupling to the environment.
    
    \item \textbf{Sideband cooling of the mechanical mode.} An in-plane magnetic field \(B_{\parallel}=60\)~mT is applied, yielding a longitudinal coupling \(g_z/2\pi \approx 180\)~kHz (see Fig.~\ref{fig:fig2}). The qubit is driven with a red-detuned tone at frequency \(\omega_r = \omega_q - \sqrt{\omega_m^2-\epsilon_r^2}\) with \(\epsilon_r = 0.89\,\omega_m\). This drive activates sideband cooling: the qubit acts as a cold reservoir for the mechanical resonator. The conditions for adiabatic elimination \(g_z \ll \Gamma\) (with \(\Gamma/2\pi = 2\)~MHz) and \(n_m\gamma_m \ll \Gamma\) (\(\gamma_m/2\pi = 5\)~Hz) are well satisfied. After cooling time of approximately 5 $\mu s$, the mechanical mode reaches its ground state with \(n_m^{\text{final}} \ll 1\). The drive is then switched off and the qubit relaxes to \(|g\rangle\). At this stage the joint state is \(|g,0\rangle\).
    
    \item \textbf{Rapid flux sweep to the avoided crossing.} The system is prepared in \(|e,0\rangle\) at \(\Phi_e = 0.3\Phi_0\) by using a $\pi$-pulse. The external flux is then swept from \(\Phi_e = 0.3\,\Phi_0\) to \(\Phi_e \approx 0.4994\,\Phi_0\) in about \(1\,\mu\text{s}\). During this sweep, the system remains predominantly in the \(|e,0\rangle\) state.
    
    \item \textbf{Coherent interaction and entanglement generation.} The system is now resonant with transverse coupling \(g_x/2\pi = \phi_{eg}\,g_\Phi/2\pi \approx 1.8\times60\)~kHz. It evolves under the Jaynes–Cummings Hamiltonian \(\hbar g_x(\hat{\sigma}_-\hat{b}^\dagger + \hat{b}\hat{\sigma}_+)\) for a variable time \(t\), about 7$\mu s$ .This produces Rabi oscillations between \(|e,0\rangle\) and \(|g,1\rangle\). At times \(t = \pi/(4g_x/2\pi), 3\pi/(4g_x/2\pi), \dots\) the system reaches maximally entangled Bell states \(\frac{1}{\sqrt{2}}(|g,1\rangle \pm i|e,0\rangle)\). The entanglement is quantified by the logarithmic negativity \(E_N(\rho)\) as shown in Fig.~\ref{fig:Entanglement}.
    
\end{enumerate}

The same in-plane field \(B_{\parallel}=60\)~mT is used throughout; if needed, \(B_{\parallel}\) can be adjusted to satisfy the cooling condition \(g_z \ll \Gamma\) while still providing a sufficiently large transverse coupling \(g_x\) at the avoided crossing. The chosen parameters are within reach of current experiments \cite{zhang_universal_2021,lee_strong_2023,gerashchenko_probing_2025}.
%

\begin{figure}[ht]
\centering
\includegraphics[width=0.48\textwidth]{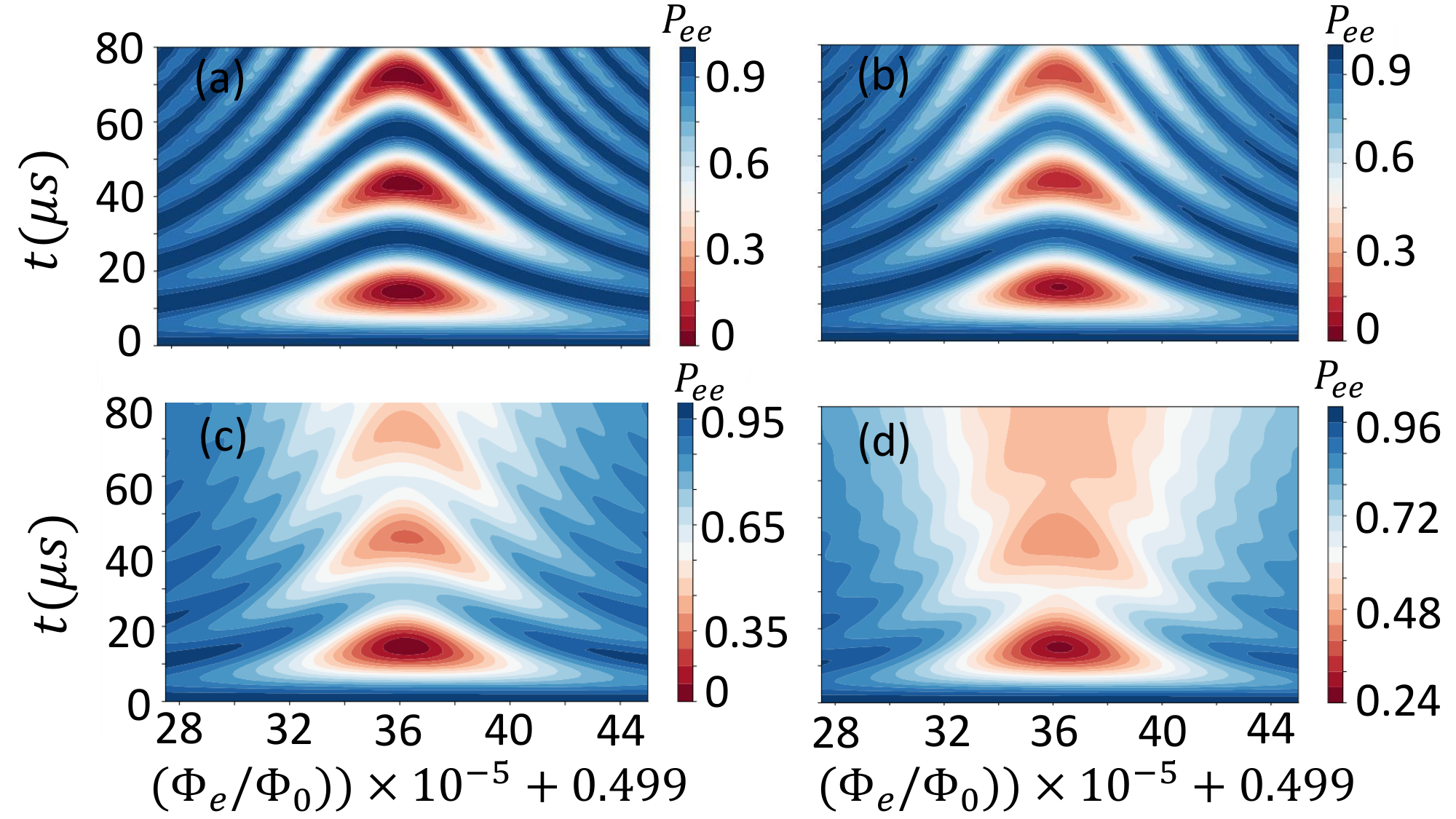}
\caption{Coherent single excitation exchange between the qubit and 
the mechanical resonator. (a) Ideal case with no energy relaxation. 
(b), (c), and (d) correspond to the energy relaxation rates of 
$\Gamma/2\pi=\Gamma_\phi/2\pi = 0.1$~kHz, $0.5$~kHz, and $1$~kHz, 
respectively. Other parameters: $n_q=n_m=50$, 
$\gamma_m/2\pi=5$~Hz, $\omega_m/2\pi=6$~MHz, 
$\Omega_d/2\pi = 10$~kHz. The increasing dephasing rates $\Gamma_\phi$ in panels (b)–(d) model the effect of
realistic flux-induced dephasing, illustrating the robustness of the excitation
exchange against experimentally relevant flux noise.The plots are generated by varying the external flux to tune the qubit
transition frequency, with the qubit initially prepared in its excited state at the avoided
crossing.
}
\label{fig:Rabi_oscillation}
\end{figure}

\begin{figure}
\centering
\includegraphics[width=0.46\textwidth]{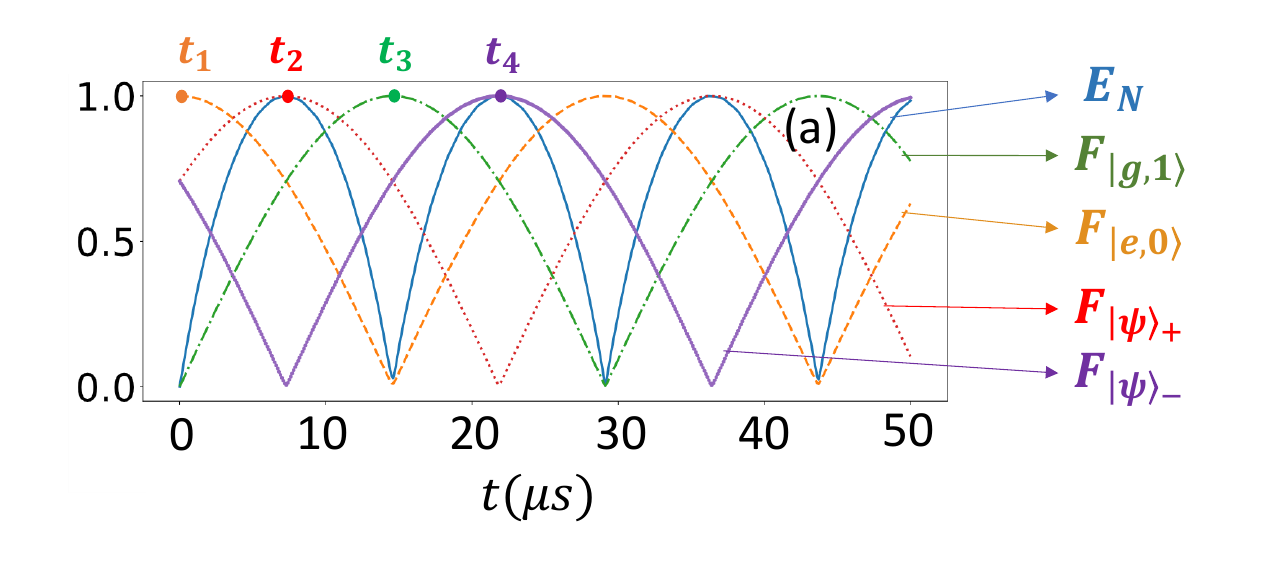}
\includegraphics[width=0.45\textwidth]{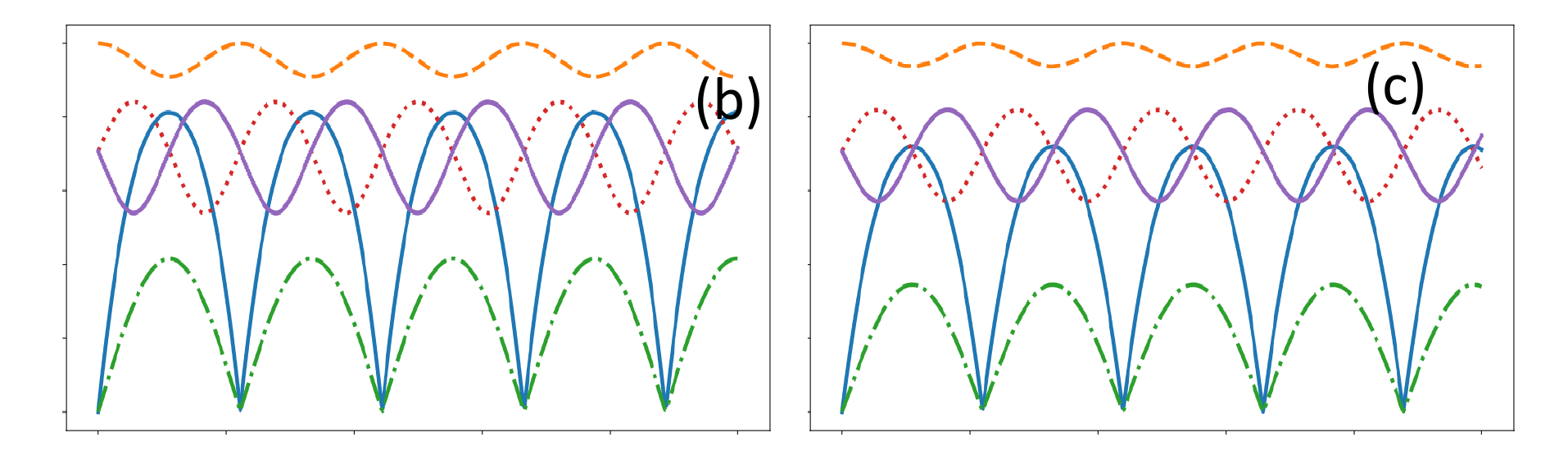}
\includegraphics[width=0.46\textwidth]{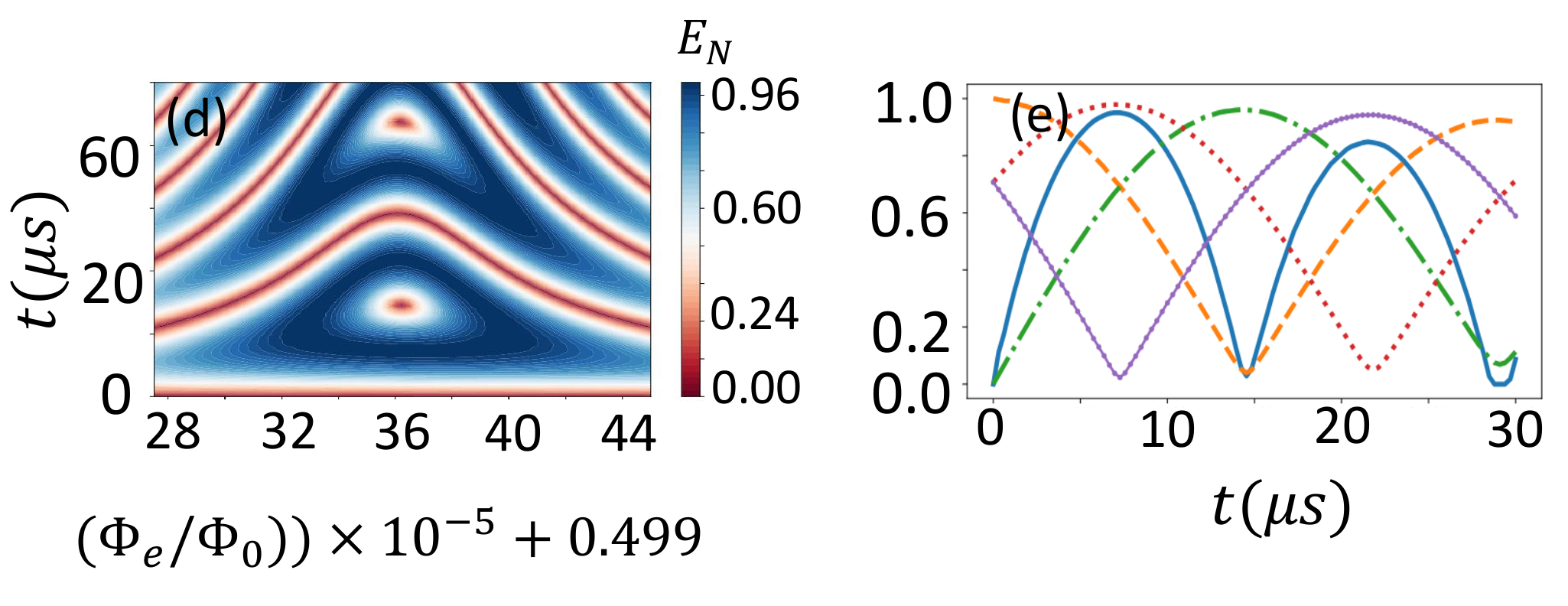}
\caption{
Plot of the entanglement measure - 
(a) Fidelities of different qubit–mechanical states \( |g,1\rangle \), \( |e,0\rangle \), and \( |\psi\rangle_\pm \), along with the entanglement 
measure \( E_N \), as the system evolves in the resonant case.  
(b) and (c) Fidelities and entanglement similar to (a), but at 
positive (b) and negative (c) detuning (\((\Delta = \omega_m - \omega_q(\Phi_e/\Phi_0) ) \)). The x- and y-axis labels in panels (b) and (c) are identical to those in panel (a).
(d) Entanglement as a function of external flux and 
evolution time.
(e) Entanglement evolution with dissipation at the flux point where qubit and mechanical resonator are in resonance.
Qubit decay rate: $\Gamma/2\pi=\Gamma_\phi/2\pi=0.1$~kHz. 
Other parameters are the same as in Fig.\ref{fig:Rabi_oscillation}}
\label{fig:Entanglement}
\end{figure}

%
%
The Rabi oscillations obtained from the qubit–mechanical Hamiltonian are shown in Fig.~\ref{fig:Rabi_oscillation}(a–d). The plots are generated for different external flux values, corresponding to different qubit transition frequencies, with the qubit initially prepared in its excited state.
Note that there will be a small decrease in the amplitude of the oscillations due to a finite contribution from diabatic transitions to the \(|g0\rangle\) state, as discussed in Appendix~\ref{appendix:C}.
To include the dissipation of the qubit and the
resonator, the system is evolved using the Lindblad 
master equation,
\begin{eqnarray}
    \label{master equation Rabi oscillation}
    \dot{\hat{\rho}} &=& -\frac{i}{\hbar}[\hat{H}+\hat{H}_d, \hat{\rho}] + [ \gamma_m(n_{m} + 1)] D[\hat{b}]\hat{\rho}
    \nonumber \\ &&+ \gamma_m n_{m} D[\hat{b}^\dagger]\hat{\rho} + [\Gamma (n_{q}+1)] D[|e\rangle\langle g|]\hat{\rho} \nonumber \\ &&
    + \Gamma n_{q}D[|e\rangle\langle g|]\hat{\rho} + \Gamma_{\Phi}D[|e\rangle\langle e|-|g\rangle\langle g|]\hat{\rho}.
\end{eqnarray}
The Rabi oscillations for different values of qubit decay 
rates are shown in Fig. \ref{fig:Rabi_oscillation}(b)-(d). 
The same Rabi oscillation can be observed at other 
avoided crossing flux points too (not shown here).

We have quantified the entanglement between the fluxonium qubit and the mechanical resonator using the logarithmic negativity $E_N(\rho)$~\cite{vidal2002computable,plenio2005introduction,plenio2005logarithmic}, defined as
\begin{equation}
E_N(\rho) = \log_2 \left[ 2 \mathcal{N}(\rho) + 1 \right],
\end{equation}
where $\mathcal{N}(\rho)$ is the negativity, given by the absolute sum of the negative eigenvalues of the partial transpose $\rho$ of the joint qubit--mechanical density matrix with respect to the qubit subsystem. This is a standard, computable entanglement monotone for mixed bipartite systems, and has been extensively used to characterize entanglement in hybrid opto- and electromechanical platforms~\cite{vitali2007optomechanical, wang2013reservoir, aspelmeyer_cavity_2014}.

In Fig.~\ref{fig:Entanglement} (a), the entanglement 
measurement in the absence of dissipation is shown at 
the qubit-mechanical resonant flux point.
As seen from the figure, the entanglement is minimal 
when the state of the qubit-mechanical system is 
either \( |g,1\rangle \) or \( |e,0\rangle \), 
as indicated by their fidelities \( F_{|g,1\rangle} \) 
and \( F_{|e,0\rangle} \) in the figure. 
This is because both \( |g,1\rangle \) and \( |e,0\rangle \) 
are separable states. 
The entanglement is maximal when the state is in the 
qubit-mechanical Bell state $|\psi\rangle_\pm = \frac{1}{\sqrt{2}} \left( |g,1\rangle \pm i\,|e,0\rangle \right).$
At times \( t_1 \), \( t_2 \), \( t_3 \), \( t_4 \) and \( t_5 \), 
the qubit-mechanical state evolves as \( |e,0\rangle \), \( |\psi\rangle_+ \), \( |g,1\rangle \),\( |\psi\rangle_- \) and \( |e,0\rangle \), respectively. 
This evolution corresponds to one complete Rabi oscillation. 
In Figs.~\ref{fig:Entanglement}(b) and \ref{fig:Entanglement}(c), 
similar plots are shown when the qubit and mechanical 
frequencies are detuned from each other. 
Fig.~\ref{fig:Entanglement}(b/c) corresponds to positive/negative detuning 
\((\Delta = \omega_m - \omega_q(\Phi_e/\Phi_0) ) \).
In both cases, the amplitude of the Rabi oscillation is not complete, and 
hence the entanglement does not reach its maximum value. 
The entanglement evolution at different flux points is shown in Fig.~\ref{fig:Entanglement}(d) for the case without dissipation. The corresponding result with dissipation is shown in Fig.~\ref{fig:Entanglement}(e), at the flux point where the qubit is resonant with the resonator.

\section{Conclusion}
\label{sec:conclusion}
To summarize, we have proposed a hybrid electromechanical 
platform based on a heavy-fluxonium qubit coupled 
to a suspended mechanical resonator via motional flux. 
By tuning the external flux bias, we accessed both 
longitudinal and transverse single-phonon coupling regimes. 
In the semiclassical regime, we observed features such 
as EIT-style transparency and mode splitting, highlighting 
the rich interaction dynamics of the system. 
The qubit's broad frequency tunability from the GHz to 
MHz range enabled sideband cooling of the mechanical resonator. 
After cooling the system, we observed coherent interaction 
in the resonant regime as evidenced by Rabi oscillations 
and entanglement near the flux-frustration point. 
Crucially, this work demonstrates that fluxonium enables 
a qualitatively different hybrid regime compared to earlier 
flux-qubit or transmon-based electromechanical systems. 
By exploiting in situ tunability of the coupling geometry 
(longitudinal \textit{vs.}\ transverse via external flux control), 
combined with long coherence at MHz operation and 
engineered dissipation for cooling, fluxonium offers capabilities 
not previously available in such platforms.
These results establish the fluxonium–mechanical 
architecture as a promising platform for quantum 
sensing, coherent phonon manipulation, and hybrid 
quantum information processing, offering both strong 
coupling and operation near flux sweet spots.

\section*{Acknowledgement}
This work is supported by MoE, Government of India (Grant No. MoE-STARS/STARS-2/2023-0161).
A.~R. acknowledges the support from UGC, GoI and Parimana fellowship by QMET Tech, DST,GoI. R.N. and A.R. contributed equally to this work.

\newpage

\bibliography{ref}

\newpage
\begin{widetext}
\appendix

\section{Pump-probe analysis}
\label{appendix A}

The Hamiltonian of the qubit-mechanical system in the 
presence of pump and probe drives at the flux point 
$\Phi_e = 0.3\,\Phi_0$ is given by Eq. \eqref{longitudinal1}. 
The equation of motion (EOM) of the system can be obtained as,
\begin{subequations}
    \label{eqn: EOM}
    \begin{eqnarray}
        \langle \dot{\hat{Q}}\rangle &=&  \omega_m\,\langle\hat{P}\rangle, \\ 
        \langle \dot{\hat{P}}\rangle &=& -\omega_m\,\langle\hat{Q}\rangle - \gamma_m\langle\hat{P}\rangle - \frac{g_{\Phi}\,\phi_{gg}}{\sqrt{2}}\langle(1-\hat{\sigma}_z )\rangle \nonumber\\ &&-\frac{g_{\Phi}\,\phi_{ee}}{\sqrt{2}}\langle(1+\hat{\sigma}_z )\rangle,   \\ 
        \langle \dot{\hat{\sigma}}_-\rangle &=& (i\Delta_q - \Gamma_q/2)\langle\hat{\sigma}_-\rangle +ig_{\Phi}(\phi_{gg}-\phi_{ee})\langle\hat{\sigma}_-(\hat{b}+\hat{b}^\dagger)\rangle  \nonumber\\ && +i\,\epsilon_d\langle\hat{\sigma}_z\rangle  + i\,\epsilon_p\langle\hat{\sigma}_z\rangle  \, e^{-i\delta_pt},\\
        \langle \dot{\hat{\sigma}}_+\rangle &=& (-i\Delta_q - \Gamma_q/2)\langle\hat{\sigma}_+\rangle -ig_{\Phi}(\phi_{gg}-\phi_{ee})\langle\hat{\sigma}_+(\hat{b}+\hat{b}^\dagger)\rangle  \nonumber\\ && - i\,\epsilon_d\langle\hat{\sigma}_z\rangle  - i\,\epsilon_p\langle\hat{\sigma}_z\rangle  \, e^{-i\delta_pt},\\      
        \langle \dot{\hat{\sigma}}_z\rangle &=& \Gamma(2n_q+1)\langle\hat{\sigma}_z\rangle - \Gamma-2i\epsilon_d\langle\hat{\sigma}_+\rangle+ 2i\epsilon_d\langle\hat{\sigma}_-\rangle  \nonumber\\ && -2i\epsilon_pe^{-i\delta_pt}\langle\hat{\sigma}_+\rangle +2i\epsilon_pe^{i\delta_pt}\langle\hat{\sigma}_-\rangle.
    \end{eqnarray}
\end{subequations}
The quantity $\Gamma_q=\Gamma(2n_q+1)+\Gamma_\Phi$ is the total 
energy decay rate of the qubit. 
The dimensionless position and momentum operators of the resonator
are defined as $\langle\hat{Q}\rangle=(\hat{b}+\hat{b}^\dagger)/\sqrt{2} $ and $\langle\hat{P}\rangle = (\hat{b}-\hat{b}^\dagger)/i\sqrt{2}$.
The mechanical resonator, due to its low frequency, 
is assumed to be in thermal state, and can be treated as a 
classical resonator coupled to the qubit. 
This leads to the semi-classical approximation 
$\langle\hat{\sigma}_\pm(\hat{b}+\hat{b}^\dagger)\rangle = \langle\hat{\sigma}_\pm\rangle\langle(\hat{b}+\hat{b}^\dagger)\rangle$. 
By substituting this in the above equation, we obtain 
the semi-classical EOM for the qubit-mechanical system. 
Taking the ansatz \( \langle\hat{O}\rangle = \langle\hat{O}_0\rangle + \langle\hat{O}_-\rangle e^{-i \delta_p t} + \langle\hat{O}_+\rangle e^{i \delta_p t} \), 
where $\hat{O}\in\{\hat{Q}, \hat{P},\hat{\sigma}_-,\hat{\sigma}_+,\hat{\sigma}_z\}$ 
and substituting in Eq. \eqref{eqn: EOM}, we get 
separate EOM for $O_0$, \(O_-\), and \(O_+\). 
Therefore, in the steady state, we get the following sets 
of equations for the mean value $\langle\hat{O_0}\rangle$
\begin{subequations}
    \label{eqn:steady state mean O_0}
    \begin{eqnarray}
        \omega_m\langle \hat{P}_0\rangle &=& 0, \\
        \langle \hat{Q}_0\rangle &=& \frac{\phi_+-\phi_-\langle\hat{\sigma}_z^0\rangle}{2\,\omega_m}, \\
        (i\Delta_q - \Gamma_q/2)\langle\hat{\sigma}_-^0\rangle  &=& -i\phi_-\langle\hat{\sigma}_-^0\rangle\langle \hat{Q}_0\rangle - i\epsilon_d\langle\hat{\sigma}_z^0\rangle, \\
        (-i\Delta_q - \Gamma_q/2)\langle\hat{\sigma}_+^0\rangle  &=&+i\phi_-\langle\hat{\sigma}_+^0\rangle\langle \hat{Q}_0\rangle + i\epsilon_d\langle\hat{\sigma}_z^0\rangle, \\
        -\Gamma(2n_q+1)\langle\hat{\sigma}_z^0\rangle  &=& + \Gamma + 2i\epsilon_d\langle\hat{\sigma}_+^0\rangle -2i\epsilon_d\langle\hat{\sigma}_-^0\rangle,
    \end{eqnarray}
\end{subequations}
which can be rearranged into 
\begin{subequations}
    \label{eqn:steady state mean rearranged}
    \begin{eqnarray}
        \langle \hat{\sigma}_z^0\rangle &=& \frac{\Gamma \eta}{-2\epsilon^2_d\Gamma_q-\Gamma(2n_q+1)\eta}, \\
        \langle \hat{\sigma}_-^0\rangle &=& \frac{-i\epsilon_d\Gamma [\Gamma_q/2+i (\Delta_q+\phi_-\langle\hat{Q}_0\rangle)]}{2\epsilon^2_d\Gamma_q+\Gamma(2n_q+1)\eta}, \\
        \langle \hat{\sigma}_+^0\rangle &=& \frac{i\epsilon_d\Gamma [\Gamma_q/2-i (\Delta_q+\phi_-\langle\hat{Q}_0\rangle)]}{2\epsilon^2_d\Gamma_q+\Gamma(2n_q+1)\eta}, \\
        \langle \hat{Q}_0\rangle &=& \frac{\phi_+-\phi_-\langle\hat{\sigma}_z^0\rangle}{2\omega_m},
    \end{eqnarray}
\end{subequations}
where $\eta=[ (\Delta_q+\phi_-\langle\hat{Q}_0\rangle)^2\,+\,\Gamma_q^2/4]$, $\phi_-=\sqrt{2}\,g(\phi_{gg}-\phi_{ee})$ and $\phi_+=\sqrt{2}\,g(\phi_{gg}+\phi_{ee})$. 
The steady state EOM of $\langle\hat{O}_-\rangle$ can be obtained by equating the coefficients of $e^{-i\delta_pt}$.
\begin{subequations}
    \label{eqn:steady state mean O-}
    \begin{eqnarray}
        -i\delta_p\langle \hat{Q}_-\rangle &=& \omega_m\langle\hat{P}_-\rangle, \\
        -i\delta_p\langle \hat{P}_-\rangle &=& -\omega_m\,\langle\hat{Q}\rangle - \gamma_m\langle\hat{P}\rangle + \phi_-\langle\hat{\sigma}_z^- \rangle \\
        -i\delta_p\langle\hat{\sigma}_-^-\rangle  &=& (i\Delta_q - \Gamma_q/2)\langle\hat{\sigma}_-^-\rangle +i\phi_-\langle\hat{\sigma}_-^-\rangle\langle\hat{Q}_0\rangle  \nonumber\\ && +i\phi_-\langle\hat{\sigma}_-^0\rangle\langle\hat{Q}_-\rangle+i\,\epsilon_d\langle\hat{\sigma}_z^-\rangle   \\
        -i\delta_p\langle\hat{\sigma}_+^-\rangle  &=& (-i\Delta_q - \Gamma_q/2)\langle\hat{\sigma}_+^-\rangle -i\phi_-\langle\hat{\sigma}_+^-\rangle\langle\hat{Q}_0\rangle  \nonumber\\ && -i\phi_-\langle\hat{\sigma}_+^0\rangle\langle\hat{Q}_-\rangle-i\,\epsilon_d\langle\hat{\sigma}_z^-\rangle  \\
        -i\delta_p\langle\hat{\sigma}_z^-\rangle  &=& -\Gamma(2n_q+1)\langle\hat{\sigma}_z^-\rangle -2i\,\epsilon_d\langle\hat{\sigma}_+^-\rangle \nonumber\\ && -2i\,\epsilon_p\langle\hat{\sigma}_+^0\rangle + 2i\epsilon_d\langle\hat{\sigma}_-^-\rangle
    \end{eqnarray}
\end{subequations}
Using Eq.~\eqref{eqn:steady state mean rearranged} and Eq.~\eqref{eqn:steady state mean O-}, we plot for the absolute value of the real part of amplitude $\langle\hat{\sigma}_-^-\rangle$ in Fig.~\ref{classical_mode split}. Here, we assume that $n_q = 0$ and $\epsilon_d, \epsilon_p \ll \Gamma_q, \Gamma$. This leads to the approximation $\langle\hat{\sigma}_z^0\rangle \approx -1$ and $\Delta_q + \phi_- \langle\hat{Q}_0\rangle \approx \Delta_q = -\omega_m$ (red detuned drive on the qubit).

\section{Higher excitation resonant and dispersive shift}
\label{Appendix B}

The higher excitation phonon-number-dependent avoided 
crossing, observed when the qubit and the resonator 
are in resonance, is shown in Fig.~\ref{fig: dispersive_shift}(a) 
and (b). 
The observation of this phonon-number-dependent shift 
is a clear signature of quantum behavior.
The shifts in the qubit and mechanical resonator 
frequencies due to detuning are shown in Fig.~\ref{fig: dispersive_shift}(c) 
and (d). When the detuning \( \Delta = (E_e - E_g)/\hbar - \omega_m \) 
is positive, the mechanical frequency shifts downward 
while the qubit frequency shifts upward. 
Conversely, when the detuning is negative, the 
mechanical frequency shifts upward and the qubit 
frequency shifts downward.
The direction of these shifts is determined by 
both the qubit state and the sign of the detuning. 
For positive detuning, the mechanical shift is 
negative when the qubit is in the ground state and 
positive when the qubit is in the excited state. 
For negative detuning, the shift direction is reversed.
This shift pattern is a typical behavior of the dispersive 
shift observed in circuit quantum electrodynamics 
(cQED) and circuit quantum acoustodynamics (cQAD) 
analysis \cite{Y_chu_2017}. 
In Fig.~\ref{fig: dispersive_shift}(c) and (d), the 
dotted green lines and solid red lines represent the 
frequency shifts corresponding to single- and double-phonon 
excitations, respectively. 
As shown, the shift due to double-phonon excitation 
is larger than that of the single-phonon case.
Recently, this characteristic behavior has been 
utilized to realize a qubit based on a mechanical 
resonator operating at GHz frequencies \cite{Yang_Yu_2024}.

\begin{figure}[ht]
\centering
\includegraphics[width=0.48\textwidth]{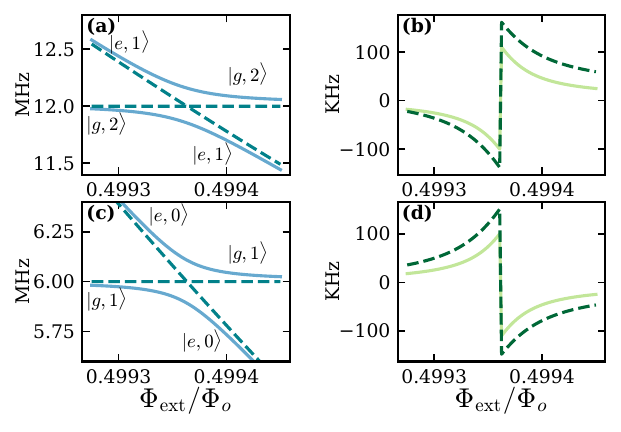}
\caption{(a) and (c): Phonon-number-dependent resonant splitting 
of the qubit–mechanical spectrum. The dotted lines 
represent the energy spectrum in the absence of coupling, 
while the solid lines show the spectrum with coupling. 
(b) and (d): Frequency shifts of the mechanical resonator 
and the fluxonium qubit due to detuning. 
The dotted lines correspond to the two-phonon excitation,
and the curves represent the shift for single-phonon excitation. 
The parameters are the same as in Fig.~\ref{fig:qubit_mech_spectrum}
}
\label{fig: dispersive_shift}
\end{figure}

\section{Landau--Zener Transition During the Sweep}
\label{appendix:C}

\begin{figure}[ht]
\centering
\includegraphics[width=0.4\textwidth]{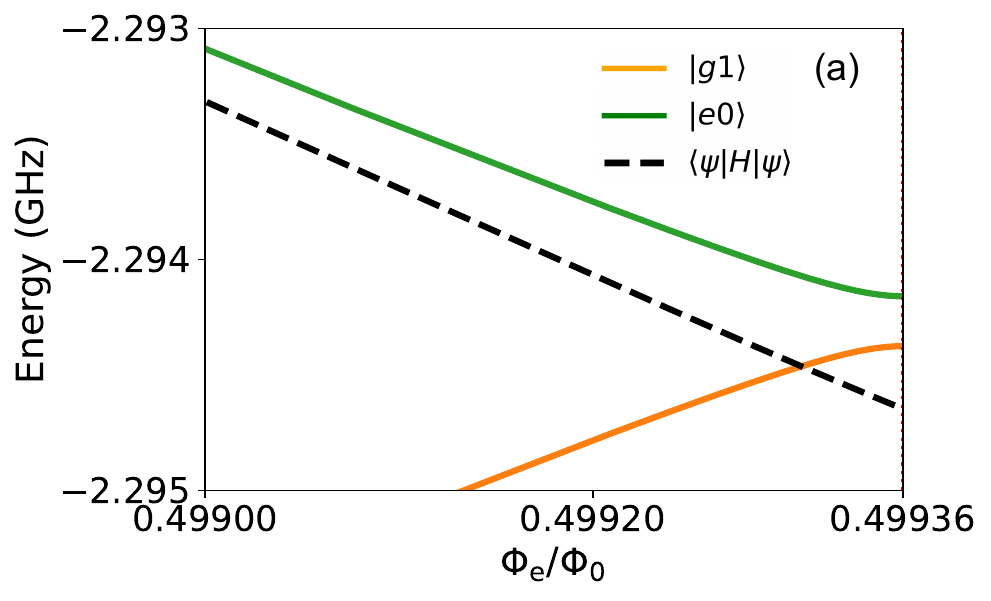}
\includegraphics[width=0.4\textwidth]{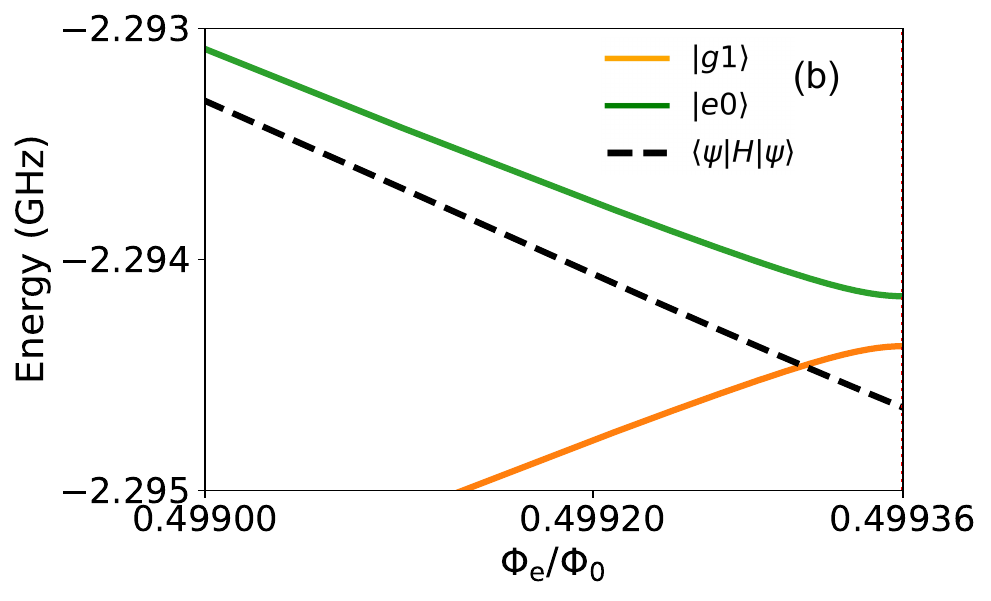}

\includegraphics[width=0.4\textwidth]{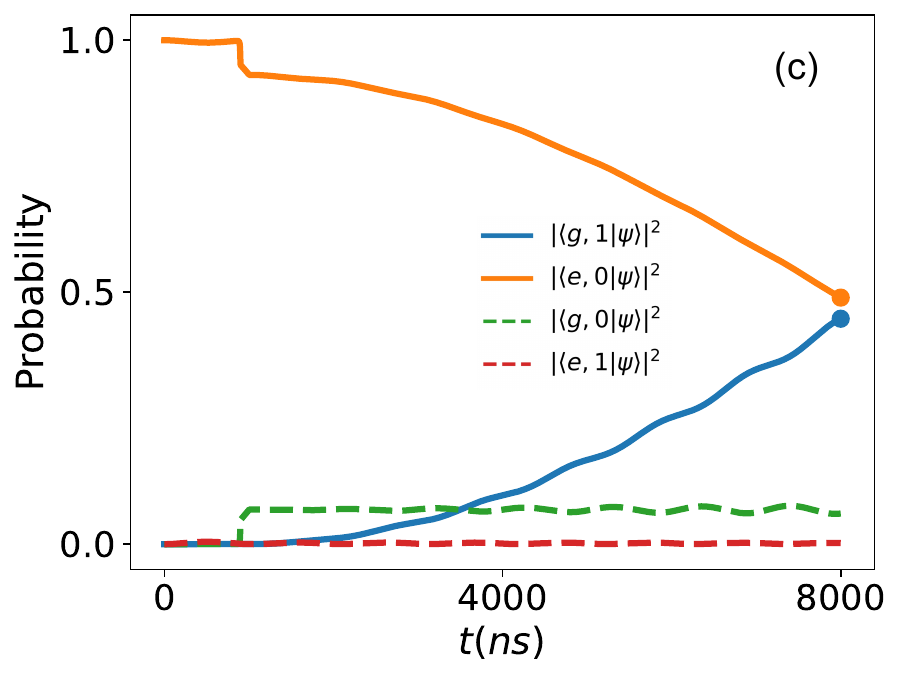}
\includegraphics[width=0.4\textwidth]{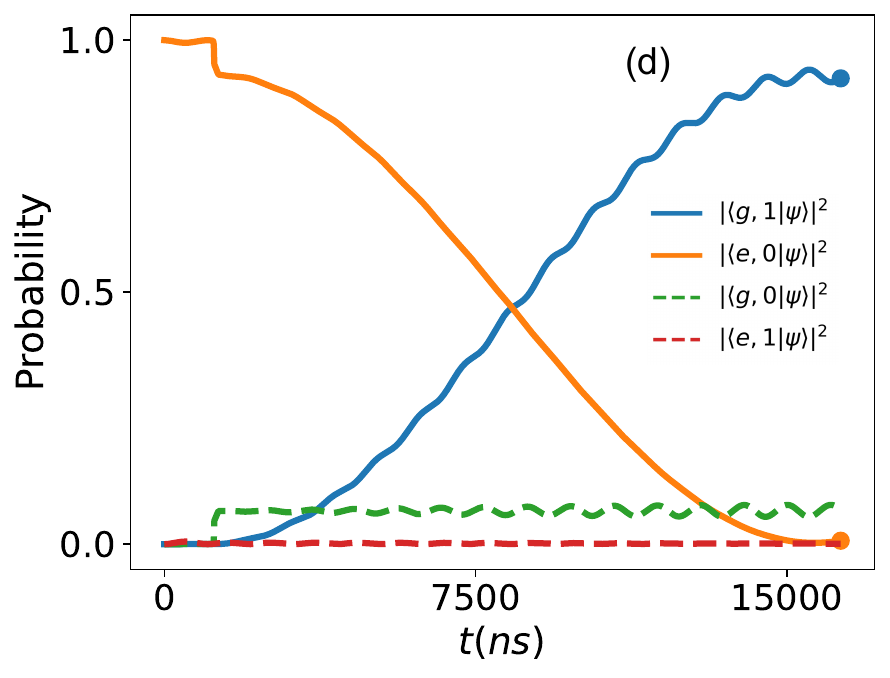}

\caption{(a) and (b) Expectation value of the evolved system (black dashed line) during a sweep from \(\Phi_e = 0.3\,\Phi_0\) to \(\Phi_e = 0.49936\,\Phi_0\), along with the instantaneous eigenvalue spectrum of the fluxonium–mechanical resonator system. The green (upper) branch corresponds to the \(|e0\rangle\) state, and the orange (lower) branch corresponds to the \(|g1\rangle\) state. The figure shows only the flux region close to the avoided crossing, and the ground-state energy is not subtracted.
The initial state of the sweep is \(|e0\rangle\) at \(\Phi_e = 0.3\,\Phi_0\). 
(c) and (d) Population dynamics of the states \(|g0\rangle\) (top dashed green line), \(|g1\rangle\) (bottom solid blue line), \(|e0\rangle\) (top solid orange line), and \(|e1\rangle\) (bottom dashed red line) during the sweep and subsequent evolution at the avoided crossing. In (c), the evolution at the crossing is shown up to the midpoint of the Rabi oscillation, while in (d) the excitation is completely transferred. Due to the Landau–Zener transition, part of the population initially in the \(|e0\rangle\) state is transferred to the \(|g0\rangle\) state before the onset of oscillations, effectively reducing the Rabi amplitude.}

\label{fig:LZtransition}
\end{figure}

To observe the coherent excitation exchange between the qubit and the resonator, we consider adiabatic sweep of the flux from the flux point at $\Phi_e = 0.3 \Phi_0$ to the avoided crossing point, i.e., $\Phi_e = 0.49936 \Phi_0$. At the same time, the mechanical–qubit interaction remains non-adiabatic throughout the sweep, allowing the excitation exchange to occur.
The Hamiltonian describing this process is given by

\begin{equation}
\hat{H}_{FMa} =
4E_C \hat{n}^2
- E_J \cos \left(\hat{\phi}+\Phi_e(t)/\phi_0\right)
+ \frac{1}{2}E_L\hat{\phi} ^2
+ \frac{1}{2}E_L (\hat{\Phi}_m(t)/\phi_0) ^2
+ E_L \hat{\phi}\,\hat{\Phi}_m(t)/\phi_0.
\label{H_Fa}
\end{equation}
The first three terms correspond to the fluxonium Hamiltonian at an instantaneous flux bias \( \phi_e(t) \). These three terms are described by the instantaneous eigenstates and eigenenergies,
\begin{equation} 
H_{Fa} = E_g(\Phi_e(t))\,|g(\Phi_e(t))\rangle\langle g(\Phi_e(t))|+ E_e(\Phi_e(t))\, e(\Phi_e(t))\rangle\langle e(\Phi_e(t))|.
\label{H_Fa_2}
\end{equation}
The last term describes the qubit–mechanical interaction.
The Hamiltonian \eqref{H_Fa} is similar to the Fluxonium-resonator Hamiltonian \eqref{Fluxonium-mech2}. 

Note that if the flux sweep were non-adiabatic, additional terms would appear in the Hamiltonian that induce transitions within the qubit and mechanical subspaces. 
Such transitions are not desirable for the process considered here.
The evolution of the population, as well as the expectation values of the energies for the states \(|g0\rangle\), \(|g1\rangle\), \(|e0\rangle\), and \(|e1\rangle\), during the sweep period of \(1.3\,\mu\text{s}\) from \(\Phi_e = 0.3\,\Phi_0\) to \(\Phi_e = 0.49936\,\Phi_0\), followed by a hold at \(\Phi_e = 0.49936\,\Phi_0\) for about 7 $\mu s$ and 14 $\mu s$ to observe the Rabi oscillations, is shown in Fig.~\ref{fig:LZtransition}. 
The initial state is \(|e0\rangle\).

As can be seen from the figure, the expectation value does not reach the middle of the avoided crossing corresponding to the state \(|e0\rangle\). 
This is due to the relatively fast sweep of \(1.3\,\mu\text{s}\) from \(\Phi_e = 0.3\,\Phi_0\) to \(\Phi_e = 0.49936\,\Phi_0\). 
As a result, part of the population is transferred to the \(|g0\rangle\) state. 
This corresponds to a diabatic transition within the intra-qubit subspace.
Due to this transition, the Rabi oscillations in the population of \(|e0\rangle\) and \(|g1\rangle\) at \(\Phi_e = 0.49936\,\Phi_0\) exhibit a reduced amplitude. If the sweep time from \(\Phi_e = 0.3\,\Phi_0\) to \(\Phi_e = 0.49936\,\Phi_0\) is increased, the amplitude of the oscillations will also increase, since the expectation value will more closely reach the center of the avoided crossing.


\end{widetext}
\end{document}